\documentstyle[aaspp4]{article}
\input epsf
\def\plotonesc#1#2{\begin{center} \leavevmode
\epsfxsize=#2\columnwidth \epsfbox{#1} \end{center}}
\def\unsetyr{\def\oyear{\relax}\def\cyear{\relax}}
\def\setyr{\def\oyear{(}\def\cyear{)}}
\unsetyr
\def\jcite#1{\setyr\cite{#1}\unsetyr}
\def\lsim{\mathrel{\hbox{\rlap{\hbox{\lower4pt\hbox{$\sim$}}}\hbox{$<$}}}}
\def\gsim{\mathrel{\hbox{\rlap{\hbox{\lower4pt\hbox{$\sim$}}}\hbox{$>$}}}}
\def\rmmat#1{{\hbox{\rm #1}}}
\def\rmscr#1{\rmmat{\scriptsize #1}}
\newcommand{\be}{\begin{equation}}
\newcommand{\ee}{\end{equation}}
\newcommand{\ba}{\begin{eqnarray}}
\newcommand{\ea}{\end{eqnarray}}
\newcommand{\ie}{{\it i.e.~}}
\newcommand{\eg}{{\it e.g.~}}
\def\d{{\rm d}}

\def\dd#1#2{\frac{\d #1}{\d #2}}

\def\eqref#1{Equation~\ref{eq:#1}}
\def\figref#1{Figure~\ref{fig:#1}}

\begin{document}
\newcommand{\bfi}{{\bf B}} \newcommand{\efi}{{\bf E}}
\newcommand{\lel}{{\lambda_e^{\!\!\!\!-}}}
\newcommand{\me}{m_e}
\newcommand{\mcs}{{m_e c^2}}
\title{The Thermal Evolution of Ultramagnetized Neutron Stars}
\author{Jeremy S. Heyl}
\authoremail{jsheyl@ucolick.org}
\author{Lars Hernquist\altaffilmark{1}}
\authoremail{lars@ucolick.org}
\affil{Lick Observatory,
University of California, Santa Cruz, California 95064, USA}
\altaffiltext{1}{Presidential Faculty Fellow}

\begin{abstract}
Using recently calculated analytic and numerical models for the thermal
structure of ultramagnetized neutron stars, we estimate the effects that
ultrastrong magnetic fields $B \ge\ 10^{14}$ G have on the thermal
evolution of a neutron star.  Understanding this evolution is necessary
to interpret models that invoke ``magnetars'' to account for soft
$\gamma$-ray emission from some repeating sources. 
\end{abstract}
\keywords{stars: neutron --- stars: magnetic fields --- radiative transfer --- 
X-rays: stars }

\section{Introduction}

Neutron stars with extremely strong magnetic dipole fields ($B \gtrsim
10^{14}$ G) may form if a helical dynamo mechanism operates
efficiently during the first few seconds after gravitational
collapse (\cite{Thom93b}) or through the conventional process of flux
freezing if the progenitor star has a sufficiently intense core field.
These ``magnetars'' initially rotate with periods $P\sim 1$ ms, but would
quickly slow down due to magnetic dipole radiation and cross the
pulsar death line after about $10^5-10^6$ yr.  With their strong
magnetic fields, magnetars have been used to explain several
phenomena including gamma-ray bursts (\cite{Usov92,Dunc92}) and soft
gamma repeaters (\cite{Thom95}).

\jcite{Shib96} have recently discussed how magnetic fields $B<10^{13.5}$
G affect neutron star cooling.  (Throughout, we will use the symbol $B$
to denote the field strength at the magnetic pole.)  They find that
below $B\sim 10^{12}$ G, the magnetic field suppresses the total heat
flux radiated by a neutron star.  However, for $B\gtrsim 10^{12}$ G,
the quantization of the electron energies enhances the conductivity
along the field lines, resulting in a net increase in the heat flux.
Here, we extend these results into the ultramagnetized regime with
$B=10^{14}-10^{16}$ G.

In this {\it Letter}, we will discuss the cooling evolution of neutron
stars which have not accreted significant material from their
surroundings, \ie neutron stars with iron envelopes.

\section{Model Envelopes}

\jcite{Heyl97a} have developed analytic models for ultramagnetized
neutron star envelopes and find that the transmitted flux through the
envelope is simply related to the direction and strength of the magnetic
field and to the core temperature ($T_c$).   Using these results as a 
guide, we numerically integrate several envelopes with $B=10^{14}-10^{16}$ G
for the case of parallel transport.   We will present the detailed results 
of these calculations in a future article.

At the outer boundary, we apply the photospheric condition (\eg
\cite{Kipp90}).  In the non-degenerate regime, photon conduction
dominates.  For the range of effective temperatures considered
free-free absorption is the most important source of opacity, and we
estimate the anisotropy factor due to the magnetic field using the
results of \jcite{Pavl76} and \jcite{Sila80}.  In the degenerate
regime, electrons dominate the conduction; we use the conductivities
of \jcite{Hern84a} or \jcite{Pote96b} and present results using both
these values.  In the semi-degenerate regime, both processes are
important, so we sum the two conductivities.

\jcite{Pote96b} give formulae to calculate electron conductivities in
the liquid and solid regimes for arbitrary magnetic field strengths.
The results of \jcite{Hern84a} are given for specific values of the
field strength.  We calculate the conductivities using the formalism and
assumptions outlined in \jcite{Hern84a} and extend his calculations to
stronger fields. 

The conductivities of \jcite{Hern84a} and \jcite{Pote96b} do not differ
in the physical processes considered in their calculation but in the
approximations employed.  In the liquid state the conductivities of
\jcite{Hern84a} tend to be approximately 15\% larger for the ground
Landau level and up to 40\% larger for the excited levels than those of
\jcite{Pote96b}.  In the solid state, the conductivities of
\jcite{Pote96b} exceed those of \jcite{Hern84a} by a factor of several;
therefore, these two models span much of the uncertainty in these
quantities. 

In the liquid state, the differences arise from two sources.  First,
both the fits of \jcite{Hern84a} and \jcite{Pote96b} for the function
$\phi(E)$ are inaccurate to $\sim 10\%$.  Second, \jcite{Hern84a}
assumes that electron-ion scattering is screened by the ion sphere; this
process dominates in the liquid regime.  \jcite{Pote96b} include Debye
and electron screening as well which dominate in the gaseous regime.
Their results are appropriate for both the gaseous and liquid regimes.
In the solid regime, \jcite{Hern84a} does not take the Debye-Waller
factor into account.  This factor tends to increase the conductivity
over a wide range of temperatures and densities (\cite{Itoh84,Pote96b}).


Our iron envelope models are calculated by adopting a plane-parallel,
Newtonian approximation.  \jcite{Hern85} found that using $Z=26$ and
$A=56$ throughout is sufficient to accurately model the envelope.  For
simplicity we fix $Z$ and $A$ to these values, rather than use the
equilibrium composition of \jcite{Baym71}.  In our approach, the core
temperature is a function of $F/g_s$, $B$ and $\psi$ (the angle
between the radial and field directions).  Here, $F$ is the transmitted heat
flux, $g_s$ is the surface gravity, and all of these values
are taken to be in the frame of the neutron star surface -- we have not 
applied the gravitational redshift to transform from the surface to the 
observer's frame.


For such strong fields, the models have a simple dependence on the angle
$\psi$; \ie $F/g_s \propto \cos^2\psi$
(\cite{Gree83,Page95,Shib95,Shib96,Heyl97a}) and furthermore the flux
for a fixed core temperature is approximately proportional to $B^{0.4}$.
With these two facts, we find that the average flux over the surface of
a neutron star with a dipole field configuration is 0.4765 times its
peak value at the magnetic poles.  We neglect the effects of general
relativity on the field configuration, which tend to make the field more
radial (\cite{Ginz64}), increasing the effects discussed here. 

\begin{figure} 
\plotonesc{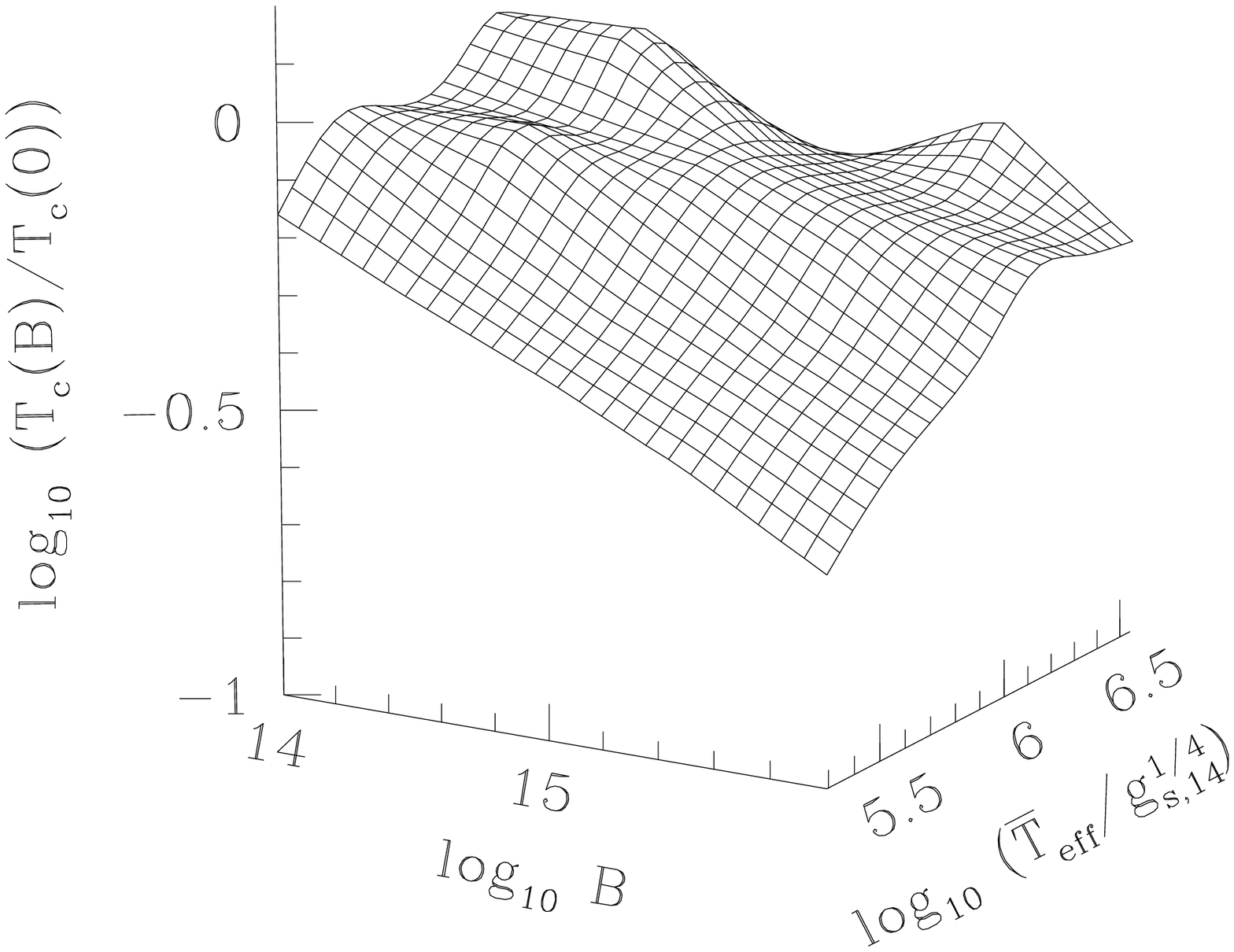}{0.5} 

\plotonesc{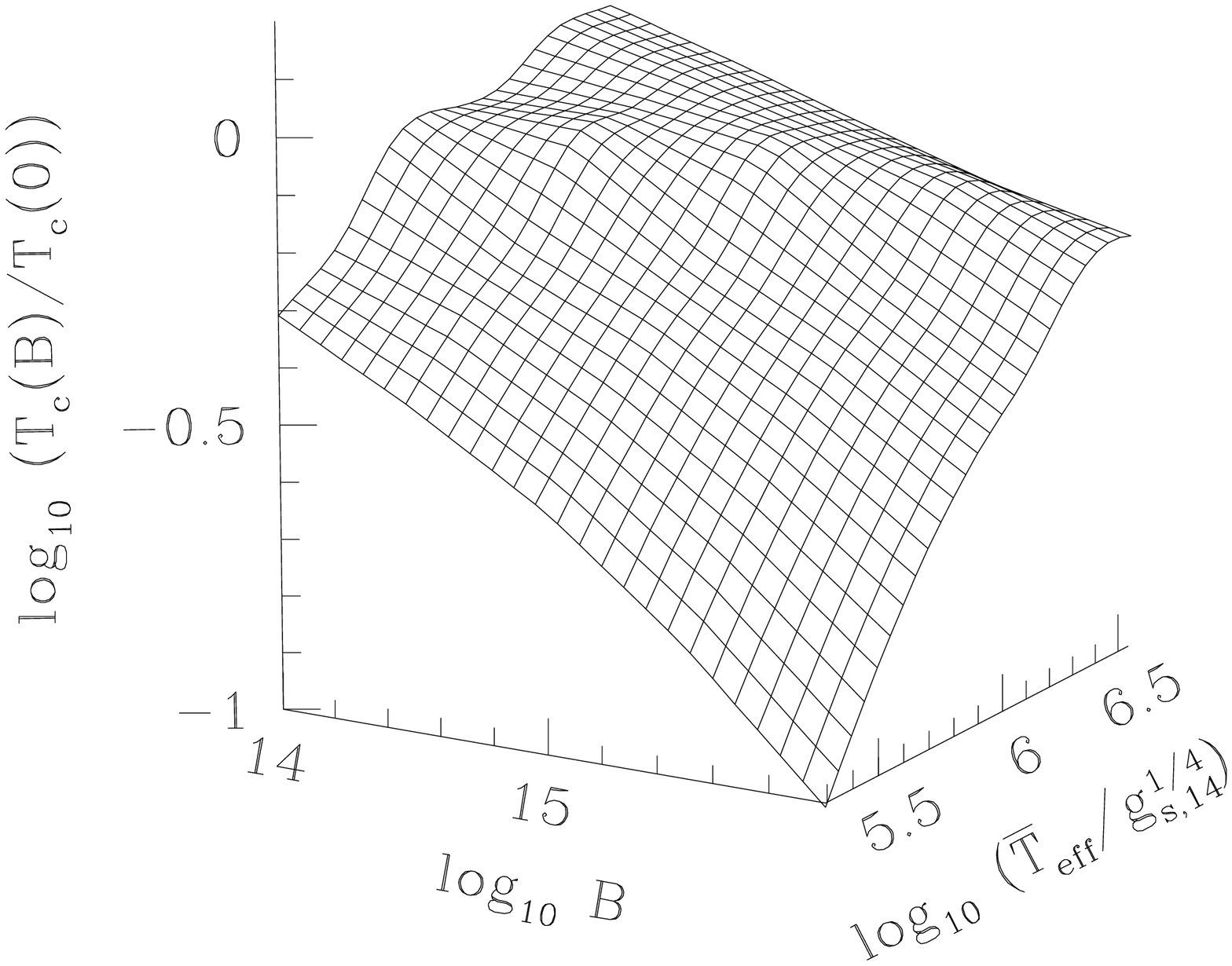}{0.5} 
\caption[]{
The upper panel depicts the ratio of the core temperature 
with $B\neq 0$ to the zero-field case (\cite{Hern84b}) using the
conductivities of \jcite{Hern84a}.  The lower panel shows the results
for the \jcite{Pote96b} conductivities.
}
\label{fig:flux3d}
\end{figure}

Using these models, we have calculated a grid of theoretical envelopes
with average effective temperatures ranging from $10^{5.4}$ K to
$10^{6.6}$ K, corresponding to a factor of $\sim 10^5$ in transmitted
flux.  We take $g_s=g_{s,14} 10^{14}$ cm/s$^2$. In the upper panel,
\figref{flux3d} depicts the ratio of the core temperature to the
zero-field case (\cite{Hern84b}) as a function of the magnetic field and
the mean effective temperature ${\bar T}_\rmscr{eff}$ over the neutron
star.  In the zero-field case, to determine the core temperature for a
given flux we combine equations~(4.7) and~(4.8) of \jcite{Hern84b},
switching from the first relation to the second when the surface
effective temperature drops below $4.25 \times 10^5$ K.  The results do
not depend qualitatively on whether equation~(4.7)~or~(4.8) of
\jcite{Hern84b} is used. The lower panel shows the ratio of the core
temperature with a magnetic field to the zero-field case for the
conductivities of \jcite{Pote96b}. 

For a given core temperature, the magnetized envelopes transmit more
heat than the unmagnetized envelopes.  For example, an effective
temperature of $3.5 \times 10^6$ K corresponds to a core temperature of
$1.1 \times 10^9$ K for an unmagnetized envelope.  With $B=10^{16}$ G,
the core temperature is $5.3 \times 10^8$ K for the \jcite{Hern84a}
conductivities and $5.8 \times 10^8$ K for the \jcite{Pote96b} ones.
Because the \jcite{Hern84a} conductivities in the liquid phase are $\sim
20$ \% larger than those of \jcite{Pote96b}, we find that the effective
temperature is slightly higher during the early cooling ($t\lsim 10^5$
yr) of the neutron star if one uses the values of \jcite{Hern84a}. 

For lower core temperatures, the insulating envelope is thinner and the
magnetic field has a stronger effect (\cite{VanR88}); the difference in
core temperatures may be even more extreme, by up to a factor of four or
ten (for \cite{Hern84a} and \cite{Pote96b} conductivities, respectively)
in the coolest envelopes considered here.  For the cooler envelopes the
relationship between the effective temperature and the core temperature
is strongly sensitive to the conductivities in the solid phase;
consequently, the Debye-Waller factor is most important during the later
cooling of the neutron star ($t \gsim 10^5$ yr).  Furthermore, during
this late phase, the partial ionization of iron may affect the equation
of state as well as the electron and photon conductivities.  This area
has not been thoroughly explored, especially at high $B$. 

For $T_c>10^8$ K, the dominant heat loss mechanism is through neutrino
emission (\eg \cite{Shap83}) which has a cooling time proportional to
$T_c^{-6}$ for the modified URCA process.  Because of this steep power
law, a factor of 1.9--2.1 difference in the core temperature (the first
example) would lead one to infer a cooling time of an ultramagnetized
neutron star 50--80 times greater than if one did not consider the
effects of the magnetic field on heat transport.  And since neutrino
cooling models generically have a cooling time proportional to
$T_c^{-\alpha}$ with $\alpha=4-6$ (\eg \cite{Shap83}), one would
generally underestimate the cooling ages of ultramagnetized neutron
stars by a large factor. 

\section{Thermal Evolution}

For $t \gtrsim 10^3$ yr, the neutron star interior has relaxed thermally
(\cite{Nomo81}), we can use the flux-to-core-temperature relation for
several values of $B$ including $B=0$ to derive the relationship between
$T_\rmscr{eff}$ and the cooling time. The technique is straightforward
during the epoch of neutrino cooling.  However, since photon emission is
enhanced, the epoch of photon cooling will begin slightly earlier, and
the time dependence of the temperature will have a slightly different
slope.  For the neutrino cooling model, we use the modified URCA process
(\eg \cite{Shap83})
\be 
L_\nu = (5.3 \times 10^{39} \rmmat{erg/s} )
\frac{M}{M_\odot} \left ( \frac{\rho_\rmscr{nuc}}{\rho} \right )^{1/3}
T_{c,9}^8 
\ee
where $T_x=T/10^x$ K. 

To understand the evolution during the photon-cooling epoch, we take
into account the surface thermal emission of photons,
\be
L_\gamma = 4 \pi R^2 {\bar T}_{\rmscr{eff}}^4
\approx 9.5 \times 10^{32} \rmmat{erg s}^{-1} 
\frac{{\bar T}_{\rmscr{eff},6}^4}{g_{s,14}} \frac{M}{M_\odot}
\ee
and we take the total thermal energy of the neutron star to be 
(\cite{Shap83})
\be
U_n \simeq 6 \times 10^{47} \rmmat{erg} \frac{M}{M_\odot} \left ( 
\frac{\rho}{\rho_\rmscr{nuc}} \right )^{-2/3} T_{c,9}^2.
\ee
Combining these equations yields 
\ba
\dd{U_n}{t} &=& - \left ( L_\nu + L_\gamma \right ) \\
\dd{T_{c,9}}{t} &=& -\frac{1}{T_{c,9}} \left [ \frac{1}{4 \times 10^7 \rmmat{yr}} 
\frac{{\bar T}_{\rmscr{eff},6}^4}{g_{s,14}} \left ( \frac{\rho}{\rho_\rmscr{nuc}}
\right )^{2/3} 
+ \frac{1}{8 \rmmat{yr}}
\left ( \frac{\rho_\rmscr{nuc}}{\rho} \right )^{1/3} T_{c,9}^8
\right ]
\ea
where $\rho$ is the mean density of the neutron star, and 
$\rho_\rmscr{nuc}=2.8 \times 10^{14}$ g cm$^{-3}$.

\begin{figure}
\plotonesc{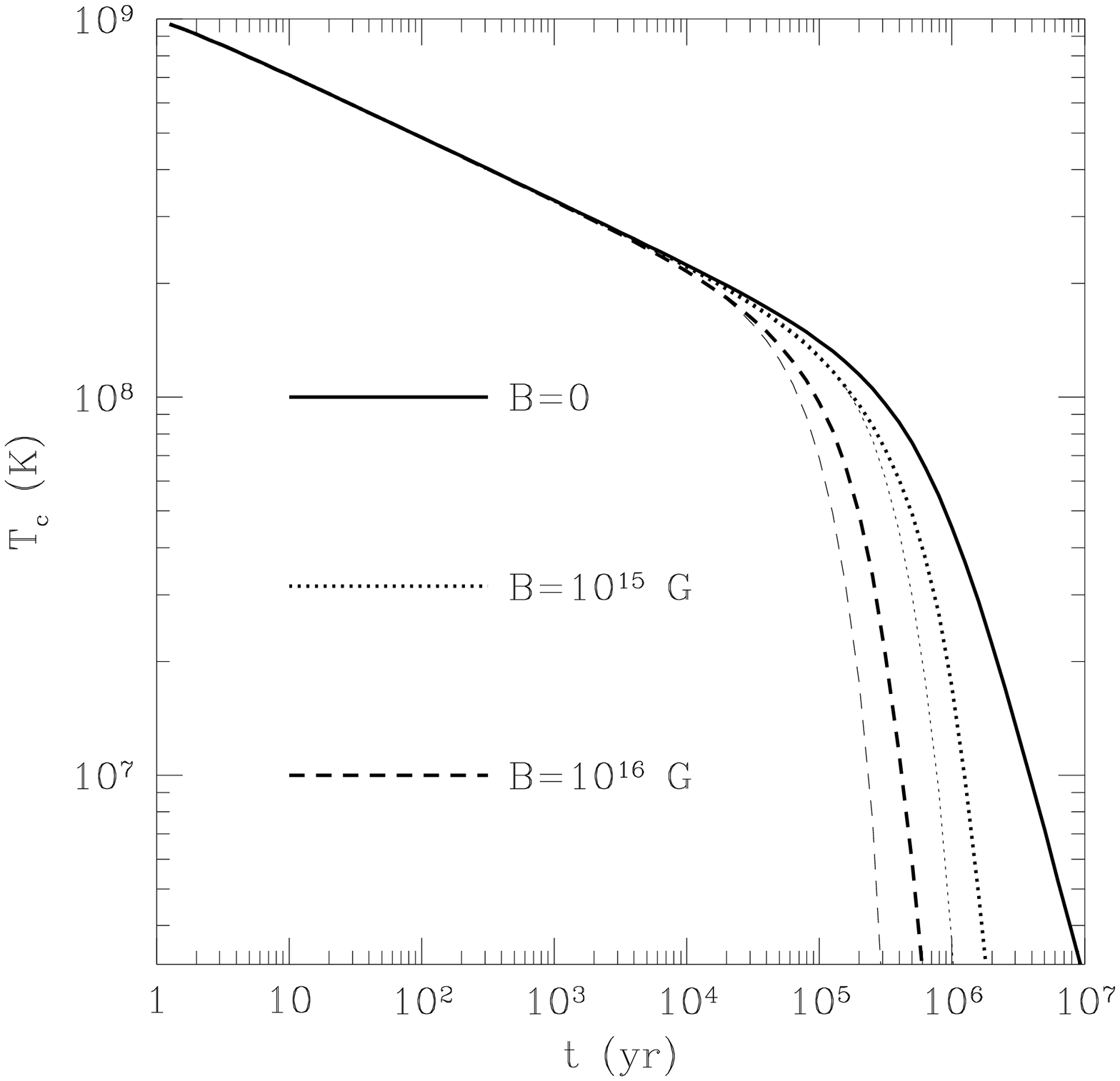}{0.5}

\plotonesc{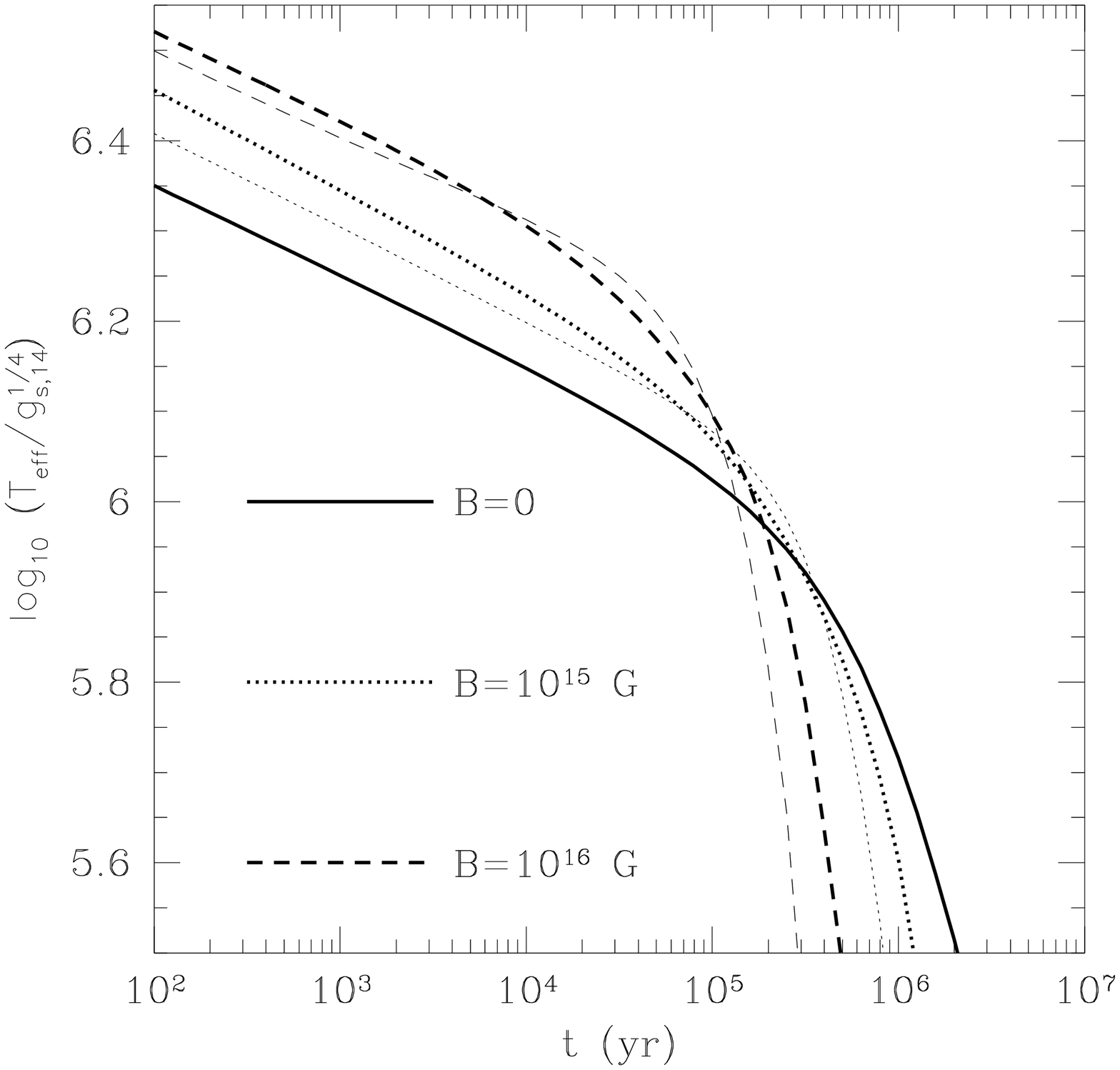}{0.5} 
\caption[]{
The upper panel depicts the evolution of the core temperature with
time for neutrino-ominated cooling (independent of $B$) and
photon-dominated cooling for several field strengths using the
conductivities of \jcite{Hern84a} (bold curves) and \jcite{Pote96b}
(light curves).  The lower panel shows the evolution of the mean
effective temperature of the cooling neutron star for the neutrino and
photon cooling epochs.  The bold and light curves designate the same
models as in the upper panel.  }
\label{fig:tevol}
\end{figure}

\figref{tevol} shows the evolution of the core temperature and mean
effective temperature at the surface.  The evolution of the core
temperature is unaffected by the magnetic field during the
neutrino-cooling epoch.  For fields approaching $10^{18}$ G, the magnetic 
field may begin to affect neutrino emission (\eg \cite{Band93}).
However, after approximately $10^6 w$ yr, photon
emission from the surface begins to dominate the evolution.  Here,
\be
w = \left ( \frac{\rho}{\rho_\rmscr{nuc}} \right )^{2/3} g_{s,14}^{-1}.
\ee

The cooling is accelerated by the magnetic field.  In the presence of
a $10^{16}$ G field, the core reaches a temperature of $10^7$ K in
only $3-6 \times 10^5 w$ yr compared to $6 \times 10^6 w$ yr for an
unmagnetized neutron star.  The bold curves trace the cooling of the
core using the \jcite{Hern84a} conductivities, and the light curves
follow the results for the \jcite{Pote96b} conductivities.

The effect is more dramatic when one compares the effective surface
temperatures of the models as a function of time.  Again we present two
sets of models.  The lower panel compares the cooling evolution using 
the conductivities of \jcite{Hern84a} (bold curves) and
using those of \jcite{Pote96b} (light curves) in the degenerate regime.

During the neutrino-cooling epoch the ultramagnetized neutron stars
($B=10^{16}$ G) have 45\%\ higher effective temperatures and emit
over four times more radiation. Because during neutrino cooling the
effective temperature falls relatively slowly with time, one can make
a large error in estimating the age of the neutron star from its
luminosity.  For example, an envelope with $10^{16}$ G field remains
above a given effective temperature 40 times longer than an
unmagnetized envelope.  For $10^{15}$ G, the timescale is increased by
up to a factor of ten.

During the photon-dominated cooling era, the enhanced flux in a strong
magnetic field reverses this effect.  Photon cooling begins to dominate
after about $10^5 w$ yr for $10^{16}$ G compared to $10^6 w$ yr in
the zero-field case.  Once photon cooling begins to dominate, the stars
with stronger magnetic fields cool more quickly.  A star with a
$10^{16}$ G field reaches a given effective temperature 3--5 times
faster than an unmagnetized star.

\section{Discussion}

Magnetic fields, especially those associated with magnetars, have a
strong effect on the observed thermal evolution of neutron stars.  
In agreement with \jcite{Shib96}, we find that during the neutrino
cooling epoch, neutron stars with strong magnetic fields are brighter
than their unmagnetized coevals.   During the photon cooling epoch, 
the situation is reversed.  A strongly magnetized neutron star cools 
more quickly during this era and emits less radiation at a given age.

It is difficult to compare our results more quantitatively with those of
\jcite{Shib96}, because besides studying more weakly magnetized neutron
stars, they make slightly different assumptions regarding the properties
of the envelope, include general relativistic effects on the magnetic
field geometry, and use the models of \jcite{VanR88} which include
Coulomb corrections to the equation of state.  For the larger fields
investigated here we do find a stronger effect than \jcite{Shib96};
however, the effect is not as strong as a naive power-law extrapolation
from $10^{13.5}$ G (the largest field studied by \cite{Shib96}) to the
ultramagnetized regime would indicate.  \jcite{Usov97} also extrapolates
results at lower field strengths and finds substantially larger photon
luminosities during the neutrino-cooling epoch than we do.  Again, a
straightforward extrapolation from weaker fields overestimates the flux
transmitted through a magnetized envelope. 

We do not find the net insulating effect that \jcite{Tsur95} find for
weaker fields of $10^{12}$ G.  \jcite{Shib96} find a similar, albeit
much weaker effect, for fields $\sim 10^{10}-10^{12}$ G.  At these field
strengths, the classical decrement in the thermal conductivity
transverse to the field direction decreases the transmitted flux for a
given core temperature.  At the much stronger fields examined here, the
increase in conductivity along the field lines (due to the quantization
of the electron energies) dominates the decrease for perpendicular
transport (in using the $\cos^2\psi$ rule we have neglected all heat
transport perpendicular to the field lines). 

\jcite{Thom95} argue that soft gamma repeaters (SGRs) are powered by
magnetic reconnection events near the surfaces of ultramagnetized
neutron stars.  Furthermore, \jcite{Ulme94} finds that a strong magnetic
field can explain the super-Eddington radiation transfer in SGRs.
\jcite{Roth94} estimate the luminosity of SGR 0526-66 in the quiescent
state to be approximately $7\times 10^{35}$ erg/sec.  Since SGR 0526-66
is located in a supernova remnant, they can also estimate the age of the
source to be approximately 5,000 years.  For an isolated neutron star
cooling by the modified URCA process, after 5,000 years, one would
expect $L_\gamma=6\times 10^{33}$ erg/s for $B=0$ and $L_\gamma=3 \times
10^{34}$ erg/s for $B=10^{16}$ G (we assume that the mass of the neutron
star is $1.4 M_\odot$).  Both these estimates fall short of the observed
value.  Even if SGR 0526-66 is powered by an ultramagnetized neutron
star, its quiescent X-ray luminosity does not originate entirely from
the thermal emission from the surface of the neutron star, unless either
the age or luminosity estimates are in error by an order of magnitude,
or possibly it has an accreted envelope.

\section{Conclusions}

We extend the previous studies of neutron star cooling into the
ultramagnetized or magnetar regime ($B\sim10^{15}-10^{16}$ G) for iron
envelopes.  We find that such an intense magnetic field dramatically
affects the thermal evolution of a neutron star.  In the
neutrino-cooling epoch, effective temperatures of ultramagnetized
neutron stars are up to 40\% larger than their unmagnetized coevals.
If the nucleons in the neutron star core are superfluid, neutrino
cooling is inhibited.  This will also increase the surface temperature
at a given epoch.

Furthermore, if one assumes an unmagnetized evolutionary track for an
ultramagnetized neutron star, one would overestimate its age by up to a
factor of twenty five.  During the photon-cooling epoch, the effect is
reversed.  Ultramagnetized neutron stars cool to a given effective
temperature three times faster than their unmagnetized counterparts. 

\acknowledgements

The work was supported in part by a National Science Foundation
Graduate Research Fellowship and the NSF Presidential Faculty Fellows
program.  We thank the referee D. G. Yakovlev for many helpful
suggestions and A. Y. Potekhin for providing the software to calculate
electron conductivities.

\begin{thebibliography}{}

\bibitem[\protect{Bander \& Rubinstein~\protect\oyear
  1993\protect\cyear}]{Band93}
Bander, M. \& Rubinstein, H. 1993,
\newblock {\em Phys Lett B,} {\bf 311}, 187.

\bibitem[\protect{Baym, Pethick \& Sutherland~\protect\oyear
  1971\protect\cyear}]{Baym71}
Baym, G., Pethick, C.~J. \& Sutherland, P.~G. 1971,
\newblock {\em ApJ,} {\bf 170}, 299.

\bibitem[\protect{Duncan \& Thompson~\protect\oyear
  1992\protect\cyear}]{Dunc92}
Duncan, R.~C. \& Thompson, C. 1992,
\newblock {\em ApJL,} {\bf 392}, 9.

\bibitem[\protect{Ginzburg \& Ozernoy~\protect\oyear
  1964\protect\cyear}]{Ginz64}
Ginzburg, V.~L. \& Ozernoy, L.~M. 1964,
\newblock {\em Zh. Exp. Teor. Fiz.,} {\bf 47}, 1030.

\bibitem[\protect{Greenstein \& Hartke~\protect\oyear
  1983\protect\cyear}]{Gree83}
Greenstein, G. \& Hartke, G.~J. 1983,
\newblock {\em ApJ,} {\bf 271}, 283.

\bibitem[\protect{Hernquist \& Applegate~\protect\oyear
  1984\protect\cyear}]{Hern84b}
Hernquist, L. \& Applegate, J.~H. 1984,
\newblock {\em ApJ,} {\bf 287}, 244.

\bibitem[\protect{Hernquist~\protect\oyear 1984\protect\cyear}]{Hern84a}
Hernquist, L. 1984,
\newblock {\em ApJS,} {\bf 56}, 325.

\bibitem[\protect{Hernquist~\protect\oyear 1985\protect\cyear}]{Hern85}
Hernquist, L. 1985,
\newblock {\em MNRAS,} {\bf 213}, 313.

\bibitem[\protect{Heyl \& Hernquist~\protect\oyear
  1997\protect\cyear}]{Heyl97a}
Heyl, J.~S. \& Hernquist, L.
\newblock {\em Almost Analtyic Models of Ultramagnetized Neutron Star
  Envelopes},
\newblock submitted to {\it MNRAS}

\bibitem[\protect{Itoh et~al.~\protect\oyear 1984\protect\cyear}]{Itoh84}
Itoh, N., Yasuharu, Kohyama, Matsumoto, N. \& Seki, M. 1984,
\newblock {\em ApJ,} {\bf 285}, 758.

\bibitem[\protect{Kippenhahn \& Weigert~\protect\oyear
  1990\protect\cyear}]{Kipp90}
Kippenhahn, R. \& Weigert, A. 1990,
\newblock {\em Stellar Structure and Evolution},
\newblock Springer, Berlin

\bibitem[\protect{Nomoto \& Tsuruta~\protect\oyear 1981\protect\cyear}]{Nomo81}
Nomoto, K. \& Tsuruta, S. 1981,
\newblock {\em ApJL,} {\bf 250}, 19.

\bibitem[\protect{Page~\protect\oyear 1995\protect\cyear}]{Page95}
Page, D. 1995,
\newblock {\em ApJ,} {\bf 442}, 273.

\bibitem[\protect{Pavlov \& Panov~\protect\oyear 1976\protect\cyear}]{Pavl76}
Pavlov, G.~G. \& Panov, A.~N. 1976,
\newblock {\em Sov. Phys. JETP,} {\bf 44}, 300.

\bibitem[\protect{Potekhin \& Yakovlev~\protect\oyear
  1996\protect\cyear}]{Pote96b}
Potekhin, A.~Y. \& Yakovlev, D.~G. 1996,
\newblock {\em A \& A,} {\bf 314}, 341.

\bibitem[\protect{Rothschild, Kulkarni \& Lingenfelter~\protect\oyear
  1994\protect\cyear}]{Roth94}
Rothschild, R.~E., Kulkarni, S.~R. \& Lingenfelter, R.~E. 1994,
\newblock {\em Nature,} {\bf 368}, 432.

\bibitem[\protect{Shapiro \& Teukolsky~\protect\oyear
  1983\protect\cyear}]{Shap83}
Shapiro, S.~L. \& Teukolsky, S.~A. 1983,
\newblock {\em Black Holes, White Dwarfs, and Neutron Stars},
\newblock Wiley-Interscience, New York

\bibitem[\protect{Shibanov et~al.~\protect\oyear 1995\protect\cyear}]{Shib95}
Shibanov, Y.~A., Pavlov, G.~G., Zavlin, V.~E. \& Tsuruta, S. 1995,
\newblock in H. B{\"{o}}hringer, G.~E. Morfill \& J.~E. Tr{\"{u}}mper (eds.),
  {\em Seventeenth Texas Symposium on Relativistic Astrophysics and Cosmology},
  Vol. 759 of {\em Annals of the New York Academy of Sciences}, p. 291, The New
  York Academy of Sciences, New York

\bibitem[\protect{Shibanov \& Yakovlev~\protect\oyear
  1996\protect\cyear}]{Shib96}
Shibanov, Y.~A. \& Yakovlev, D.~G. 1996,
\newblock {\em A\&A,} {\bf 309}, 171.

\bibitem[\protect{Silant'ev \& Yakovlev~\protect\oyear
  1980\protect\cyear}]{Sila80}
Silant'ev, N.~A. \& Yakovlev, D.~G. 1980,
\newblock {\em Astrophys. Sp. Sci.,} {\bf 71}, 45.

\bibitem[\protect{Thompson \& Duncan~\protect\oyear
  1993\protect\cyear}]{Thom93b}
Thompson, C. \& Duncan, R.~C. 1993,
\newblock {\em ApJ,} {\bf 408}, 194.

\bibitem[\protect{Thompson \& Duncan~\protect\oyear
  1995\protect\cyear}]{Thom95}
Thompson, C. \& Duncan, R.~C. 1995,
\newblock {\em MNRAS,} {\bf 275}, 255.

\bibitem[\protect{Tsuruta \& Qin~\protect\oyear 1995\protect\cyear}]{Tsur95}
Tsuruta, S. \& Qin, L. 1995,
\newblock in H. B{\"{o}}hringer, G.~E. Morfill \& J.~E. Tr{\"{u}}mper (eds.),
  {\em Seventeenth Texas Symposium on Relativistic Astrophysics and Cosmology},
  Vol. 759 of {\em Annals of the New York Academy of Sciences}, p. 299, The New
  York Academy of Sciences, New York

\bibitem[\protect{Ulmer~\protect\oyear 1994\protect\cyear}]{Ulme94}
Ulmer, A. 1994,
\newblock {\em ApJL,} {\bf 437}, 111.

\bibitem[\protect{Usov~\protect\oyear 1992\protect\cyear}]{Usov92}
Usov, V.~V. 1992,
\newblock {\em Nature,} {\bf 357}, 472.

\bibitem[\protect{Usov~\protect\oyear 1997\protect\cyear}]{Usov97}
Usov, V.~V. 1997,
\newblock {\em A\&A,} {\bf 317}, 87.

\bibitem[\protect{{Van Riper}~\protect\oyear 1988\protect\cyear}]{VanR88}
{Van Riper}, K.~A. 1988,
\newblock {\em ApJ,} {\bf 329}, 339.

\end{thebibliography}
\end{document}